\documentclass[aps,prb,twocolumn, superscriptaddress , amsmath, amssymb,reprint]{revtex4-1}
\usepackage{graphicx}
\usepackage{dcolumn}
\usepackage{bm}
\usepackage{upgreek}
\usepackage{times}
\usepackage{mathrsfs}
\usepackage{multirow}
\usepackage{fancyhdr} 
\usepackage{hyperref} 
\usepackage[version=3]{mhchem}

\begin{document}

\title{Ionization energy as a stability criterion for halide perovskites}

\author{Chao Zheng}
\email[E-mail:~]{zhengc8@mcmaster.ca}
\author{Oleg Rubel}
\affiliation{Department of Materials Science and Engineering, McMaster University, 1280 Main Street West,
Hamilton, Ontario L8S 4L8, Canada}

\date{\today}

\begin{abstract}
Instability of hybrid organic-inorganic halide perovskites hinders their development for photovoltaic applications. First-principle calculations are used for evaluation of a decomposition reaction enthalpy of hybrid halide perovskites, which is linked to experimentally observed degradation of device characteristics. However, simple criteria for predicting stability of halide perovskites are lacking since Goldschmidt's tolerance and octahedral geometrical factors do not fully capture formability of those perovskites. In this paper, we extend the Born-Haber cycle to partition the reaction enthalpy of various perovskite structures into lattice, ionization, and molecularization energy components. The analysis of various contributions to the reaction enthalpy points to an ionization energy of a molecule and a cage as an additional criterion for predicting chemical trends in stability of hybrid halide perovskites. Prospects of finding new perovskite structures with improved chemical stability aimed for photovoltaic applications are discussed.
\end{abstract}

\pacs{TBD}

\maketitle

%
%
\section{Introduction}
The efficiencies of hybrid organic-inorganic perovskite solar cells have already increased to over 20\%~\cite{Lee_S_338_2012, Park_JPCL_4_2013, Jung_S_4_2015, Yang_S_12_2015}. Fabrication of hybrid organic perovksite is based on a low temperature solution method, thus offering a low-cost alternative to crystalline thin-film photovoltaic devices. The main obstacle hindering the commercialization of hybrid organic perovskite solar cells is the instability of the active material. Hybrid perovskites are prone to a phase separation that takes place instantly under ambient conditions (moisture, UV radiation, atmospheric oxygen, etc.) \cite{Christians_JACS_137_2015, Wozny_CM_27_2015, Buin_CM_27_2015}. The detrimental role of moisture in creating a degradation pathway for halide perovskites was previously discussed from acid-base chemistry \cite{Frost_NL_14_2014}, molecular dynamic simulations \cite{Mosconi_CM_27_2015,Zhang_JPCC_119_2015}, hydrolysis reaction \cite{Zhao_SR_6_2016} and thermodynamic \cite{Tenuta_SR_6_2016} perspectives. Encapsulation of the perovskite cells does not prevent their degradation either. The active layer of encapsulated hybrid organic perovskites eventually decompose after a period of time that ranges from several days to a month \cite{Burschka_N_499_2013, Han_JMCA_3_2015}.

Intrinsic instability of hybrid halide perovskite structures can be captured at the level of first-principle calculations~\cite{Zhang_arXiv_2015, Ganose_JPCL_6_2015, Fedwa_SR_6_2016} by evaluating the enthalpy of the reaction 
\begin{equation}\label{Eq:Decomosition_reaction_general}
	AX + BX_2 \rightarrow ABX_3
\end{equation}
based on the total energy of the solid compounds involved. Here $A$ represents an organic cation, $B$ and $X$ are the metal and halide elements, respectively. The negative reaction enthalpy $\Delta H_\text{r}$ indicates stable products. The lower the value of $\Delta H_\text{r}$, the more stable the structure is against decomposition. For example, the reaction enthalpy for tetragonal \ce{CH3NH3PbI3} is within the range of $-0.1\ldots0.06$~eV per formula unit~\cite{Zhang_arXiv_2015,Buin_CM_27_2015,Tenuta_SR_6_2016}, which renders the structure to be at the boundary between weakly stable and unstable agreeing with experimental observations \cite{Buin_CM_27_2015}. Despite the success of first-principle calculations in predicting formability of hybrid halide perovskite structures, the origin of intrinsic instability and avenues for its improvement remain unclear.

Geometrical factors such as the Goldschmidt's tolerance factor~\cite{Goldschmidt_N_14_1926} and octahedral factor successfully explain formability of various inorganic perovskite structures~\cite{Li_JAC_372_2004}. The tolerance factor $t$ measures compactness of the perovskite structure. The value of the tolerance factor for \ce{CH3NH3PbI3} is $t=0.91$~\cite{Nagabhushana_PNAS_113_2016}, which is within the range of acceptable values $t=0.8-0.95$~\cite{Li_JAC_372_2004}. \citet{Li_JAC_372_2004} pointed out that the tolerance factor alone does not fully capture formability of perovkite structures and proposed to add Pauling's octahedral factor \cite{Pauling_JACS_51_1929} $r_B/r_X$ ($r_B$ and $r_X$ are the ionic radii of cation $B$ and anion $X$, respectively) as an additional geometrical criterion. In the case of \ce{CH3NH3PbI3} the octahedral factor $r_\text{Pb}/r_\text{I}=0.54$ is within the allowable range of $0.414-0.732$~\cite{Pauling_JACS_51_1929}. This analysis suggests that geometrical factors are not sufficient to explain the instability of hybrid halide perovskites.

\citet{Frost_NL_14_2014} attributed the instability of hybrid organic halide perovskites to a relatively low electrostatic lattice energy of their ionic structure as compared to non-halide perovskite compounds. For instance, traditional inorganic perovskites of the \ce{II-IV-VI3} family, e.g. \ce{PbTiO3}, have the lattice energy of $-119$~eV. This value is much lower that the lattice energy of $-28$~eV for \ce{CH3NH3PbI3} perovksite, which belongs to the \ce{I-II-VII3} family. This argument suggests that \ce{I-II-VII3} perovskites have intrinsically lower electrostatic energy and thus weaker chemical stability. On the other hand, the experimental reaction enthalpy for \ce{PbTiO3} is only $-0.38$~eV \cite{Rane_JSSC_161_2001}, which is orders of magnitude less than its lattice energy. It is also known that \ce{CsPbI3} perovskite structure is indeed stable up to the temperature of 460$^\circ$C \cite{Sharma_Z_175_1992}, above which the material melts without decomposition, despite of its higher lattice energy of $-27$~eV. These observations indicate that the lattice energy alone cannot be used as a criterion for stability of ionic structures.

The Born-Haber cycle is traditionally used for analysis of formation enthalpies. It allows to break the formation energy into the following components: atomization enthalpy, ionization enthalpy, and lattice enthalpy \citep{Treptow_FCE_74_1997}. In this paper we extend the Born-Haber cycle to the analysis of energy components of the reaction enthalpies for various perovskite structures using the density functional theory (DFT). It will be shown that in \ce{I-II-VII3} organic and inorganic perovskites the lattice energy contribution is largely cancelled by the molecularization energy leaving  the ionization enthalpy to determine the direction of the reaction. The  instability of hybrid organic lead-iodine perovskites can be attributed to the high energy associated with ionization of organic molecules and \ce{[PbI3]-}.

\section{Basic concepts}

The Born-Haber cycle was originally proposed by Max Born and Fritz Haber as a way to measure formation energies of ionic structures \citep{Treptow_FCE_74_1997}. The cycle also provides a method to determine the lattice energy of the structures, which otherwise cannot be directly measured experimentally. Here we will explain the essence of the Born-Haber cycle and its utilization for analysis of reaction enthalpy components using the \ce{CH3NH3PbI3} perovksite structure as an example. 

\begin{figure*}
 \centering
 \includegraphics[width=0.7\textwidth]{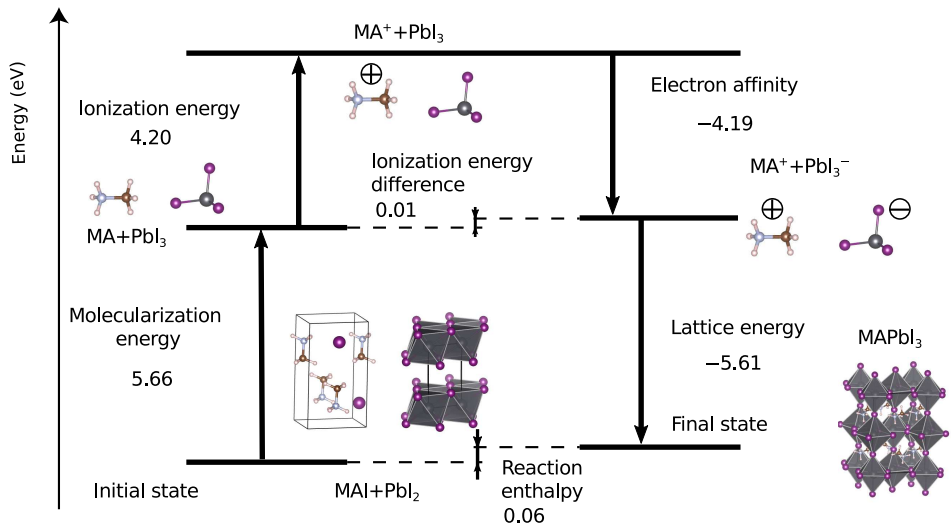}
 \caption{Born-Haber cycle of hybrid halide perovskites: Methylammonium (MA) lead iodide obtained with \ce{[CH3NH3]+} and \ce{[PbI3]-} ions as elementary species.}
 \label{fgr:BH_cycle_MAPbI3-ion}
\end{figure*}

The formation process of \ce{CH3NH3PbI3} from solid \ce{CH3NH3I} and \ce{PbI2} compounds can be subdivided into several consecutive steps illustrated in Fig.~\ref{fgr:BH_cycle_MAPbI3-ion}.

The initial step---molecularization (similar to the atomization in the original Born-Haber cycle)---involves breaking the \ce{CH3NH3I} and \ce{PbI2} lattice structures and formation of \ce{CH3NH3} and \ce{PbI3} molecules
\begin{equation}\label{Eq:BH_MAPbI3_step1}
	\text{CH}_3\text{NH}_3\text{I}(\text{s}) + \text{PbI}_2(\text{s}) 
	\xrightarrow{\Delta H_\text{mo}}
	\text{CH}_3\text{NH}_3(\text{g}) + \text{PbI}_3(\text{g})~.
\end{equation}
The rational for using \ce{CH3NH3} and \ce{PbI3} molecules as the smallest units in the Born-Haber cycle is justified by the existence of the corresponding free standing ions \cite{Cremaschi_JMS_29_1975, Lanford_JACS_63_1941}, and will be discussed in section \ref{Sec:Results}.

The next step is the ionization of \ce{CH3NH3} molecule
\begin{equation}\label{Eq:BH_MAPbI3_step2a}
	\text{CH}_3\text{NH}_3(\text{g}) + \text{PbI}_3(\text{g})
	\xrightarrow{\Delta H_\text{ion,1}}
	\text{CH}_3\text{NH}_3^+(\text{g}) + \text{PbI}_3(\text{g})~,
\end{equation}
followed by the ionization of \ce{PbI3}
\begin{equation}\label{Eq:BH_MAPbI3_step2b}
	\text{CH}_3\text{NH}_3^+(\text{g}) + \text{PbI}_3(\text{g})
	\xrightarrow{\Delta H_\text{ion,2}}
	\text{CH}_3\text{NH}_3^+(\text{g}) + \text{PbI}_3^-(\text{g})~.
\end{equation}
It can be seen from the diagram in Fig.~\ref{fgr:BH_cycle_MAPbI3-ion} that the formation of \ce{[CH3NH3]+} ion is an endothermic process, whereas the ionization of \ce{PbI3} is an exothermic process. The resultant ionization energy is an additive of two enthalphies
\begin{equation}\label{Eq:BH_MAPbI3_dHion}
	\Delta H_\text{ion} = \Delta H_\text{ion,1} + \Delta H_\text{ion,2}~.
\end{equation}

Finally, electrically charged \ce{[CH3NH3]+} and \ce{[PbI3]-} complex ions are combined to form \ce{CH3NH3PbI3} crystalline structure
\begin{equation}\label{Eq:BH_MAPbI3_step3}
	\text{CH}_3\text{NH}_3^+(\text{g}) + \text{PbI}_3^-(\text{g})
	\xrightarrow{\Delta H_\text{latt}}
	\text{CH}_3\text{NH}_3\text{PbI}_3(\text{s})~.
\end{equation}
The amount of energy $\Delta H_\text{latt}$ released in this reaction is called the lattice energy of the hybrid organic perovskite structure. This concludes the Born-Haber cycle of \ce{CH3NH3PbI3}. The total reaction enthalpy is compiled from enthalpies of individual steps of the cycle
\begin{equation}\label{Eq:BH_MAPbI3_dHr}
	\Delta H_\text{r} = \Delta H_\text{mo} + \Delta H_\text{ion} + \Delta H_\text{latt}~.
\end{equation}

 \section{Computational details}

Electronic structure calculations have been performed in the framework of DFT \cite{Kohn_PR_140_1965} and Perdew-Burke-Ernzerhof generalized gradient approximation \cite{Perdew_PRL_77_1996} (GGA-PBE) for the exchange-correlation functional. Total energies of all compounds were obtained using the Vienna \textit{ab initio} simulation program (VASP) and projector augmented-wave (PAW) potentials \cite{Kresse_PRB_54_1996,Kresse_PRB_59_1999,Blochl_PRB_50_1994}. 

All crystal structures of compounds studied here are taken at their most stable polymorph at ambient conditions. Among perovskite structures, \ce{CH3NH3PbI3} adapts a tetragonal $\beta$-phase at the ambient temperature, \ce{CH3NH3PbBr3} and \ce{CH3NH3PbCl3} have a cubic phase \cite{Poglitsch_JCP_87_1987, Onoda_JPCS_51_1990}, \ce{CN3H6PbI3}  and \ce{(CH3)4NPbI3} favor hexagonal structures \cite{Giorgi_JPCC_9_2015, Dimesso_MSEB_204_2016, Liu_IC_55_2016}. \ce{CsPbI3}, \ce{CsPbBr3}, and \ce{CsPbCl3}  prefer an orthorhombic (Pnma) $\delta$-phase \cite{Sharma_ZPC_175_1992,Hidaka_PSSA_79_1983}. The crystal structure of \ce{CH3NH3I}, \ce{CH3NH3Br}, and \ce{CH3NH3Cl} organic salts correspond to $\alpha'$-tetragonal (P4/nmm) phase at room temperature  \cite{Ishida_ZNA_50_1995,Gabe_AC_14_1961,Hughes_JACS_68_1946}. \citet{Szafranski_CEC_15_2013} reported the structures of guanidinium iodide \ce{CN3H6I}, and the structure of tetramethylammonium iodine \ce{(CH3)4NI} was obtained using \ce{(CH3)4NAu} \cite{Dietzel_CC_21_2001} as a parent structure followed by full relaxation of their structural parameters. Cubic crystal structures of CsI, CsBr, CsCl and NaCl as well as hexagonal \ce{PbI2} and orthorhombic \ce{PbBr2} were taken from \citet{Ralph_2_1960, Gerlach_ZPAHN_9_1922}. The crystal structure of orthorhombic \ce{PbCl2} was derived from the structure of \ce{PbBr2}.

For reciprocal space integration, $4\times4\times4$ Monkhorst-Pack grid \cite{Monkhorst_PRB_13_1976} was used for cubic phases, $3\times3\times2$  were used for tetragonal phases, $4\times4\times3$ for hexagonal phases and $3\times6\times2$ for orthorhombic \ce{CsPbX3} phases and $4\times8\times4$ for ohthorhombic \ce{PbBr2} and \ce{PbCl2}. The cutoff energy for a plane wave expansion was set at 400~eV. The lattice constant and atomic positions were optimized such that residual forces acting on atoms did not exceed 2~meV/{\AA}, and the residual hydrostatic pressure was less than 50~MPa.

Gaseous phases, such as Cs, \ce{[CH3NH3]+}, \ce{[PbI3]-}, were modelled as an individual atom/molecule surrounded by 20~{\AA} of vacuum. All calculations related to gaseous phases were performed in conjunction with optimization of internal degrees of freedom. Only $\Gamma$-point was used in the Brillouin zone. The ionization energy of positively charged ions was calculated by subtracting the total energy of cations (e.g. Cs$^+$, \ce{[CH3NH3]+}, \ce{[CN3H6]+}) from the energy of neutral atoms or molecules (e.g. Cs, \ce{CH3NH3}, \ce{CN3H6}). Similarly, the electron affinity of negatively charged ions was modelled by adding one electron to \ce{PbCl3}, \ce{PbBr3}, or \ce{PbI3} molecules to form \ce{[PbCl3]-}, \ce{[PbBr3]-}, and \ce{[PbI3]-} anions. The electron affinity of these ions was represented as an energy difference between negatively charged complex ions and neutral species. Monopole, dipole and quadrupole corrections implemented in VASP \cite{Makov_PRB_51_4014, Neugebauer_PRB_46_16067} were used for eliminating leading errors and acquiring accurate total energies of all charged ions.

\texttt{VESTA}~3 package \cite{Momma_JAC_44_2011} was used to visualize crystal structures and for computing the Madelung electrostatic energy using oxidation state as  formal charges. In these calculations, the radius of ionic sphere and the reciprocal-space range were set at 1~{\AA} and 4~{\AA$^{-1}$}, respectively.

\section{Results and discussion} \label{Sec:Results}

\subsection{Lattice energies of halide perovskites}

\begin{figure*}
 \centering
 \includegraphics[width=0.7\textwidth]{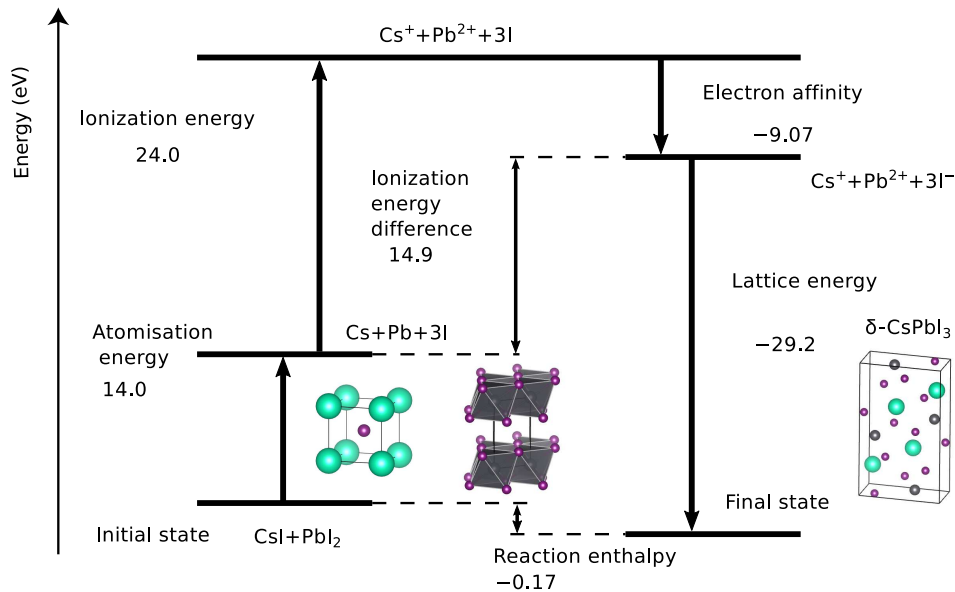}
 \caption{Born-Haber cycle of inorganic halide perovskites: Caesium lead iodide obtained with Cs$^+$, Pb$^{2+}$, and I$^-$ ions as elementary species.}
 \label{fgr:BH_cycle_CsPbI3-ion}
\end{figure*}

Calculation of individual energies associated with various steps in the Born-Haber cycle requires subdivision of the ionic solid in question into elementary species. In the case of alkali halides (such as NaCl, CsCl, etc.), the atomization is an apparent choice. Following the same strategy, \ce{Cs}$^+$, Pb$^{2+}$, and I$^-$ ions can be used to calculate the lattice energy, which yields $\Delta H_\text{latt}\simeq-29$~eV (Fig.~\ref{fgr:BH_cycle_CsPbI3-ion}).

This value agrees well with the Madelung energy of $-27$~eV obtained from the point charge model. \citet{Gopal_JICS_30_1953} noticed existence of a trend between the lattice energy $\Delta H_\text{latt}$ and the melting point $T_m$ of alkali halides with the proportionality factor of $-\Delta H_\text{latt}/T_m\approx7.4\cdot10^{-3}$~eV/K. Assuming that the same proportionality holds for perovskite structures, the melting point of \ce{I-II-VII3} perovskites would be near 3900~K, which is an order of magnitude greater than the actual values of $733-888$~K for group-I lead halide perovskites (\ce{CsPbI3}, \ce{CsPbBr3}, and \ce{CsPbCl3}) \cite{, Stoumpos_CGD_13_2013, Fayon_Ferroelectrics_185_1996}.

Alternatively, we can separate \ce{CsPbI3} perovskite structure into two ions \ce{Cs}$^+$ and \ce{[PbI3]-}. The existence of the corresponding free-standing ions was verified experimentally \cite{Cremaschi_JMS_29_1975, Lanford_JACS_63_1941}. Using this approach we re-evaluated the lattice energy of \ce{CsPbI3} as $-5.55$~eV using the Born-Haber cycle similar to that shown in Fig.~\ref{fgr:BH_cycle_MAPbI3-ion}. This result translates into a substantially lower melting point of approximately 750~K, which is remarkably close to the experimental value of 749~K.

Similar calculations of the lattice energy were performed for other inorganic \ce{I-II-VII3} and \ce{II-IV-VI3} perovksites. Results are summarized in Table~\ref{tbl:BH-perovskites}.

\begin{table*}
    \caption{Components (eV) of the reaction enthalpies extracted from Born-Haber cycle as well as the melting temperature and stability against spontaneous decomposition for halide perovskites and other ionic structures.}\label{tbl:BH-perovskites}
    \begin{ruledtabular}
        \begin{tabular}{lrrrrlc}
    Compounds & $\Delta H_\text{mo}$ & $\Delta H_\text{latt}$ & $\Delta H_\text{ion}$ & $\Delta H_\text{r}$ & $T_\text{m}$~(K) & Stability \\
    \hline
    $\delta$-\ce{CsPbCl3} & 6.23 & $-$5.95 & $-$0.67 & $-$0.39 & 888 \cite{Fayon_Ferroelectrics_185_1996} & Y \\
    $\delta$-\ce{CsPbBr3} & 5.80 & $-$5.72 & $-$0.46 & $-$0.39 & 840 \cite{Stoumpos_CGD_13_2013} & Y \\
    $\delta$-\ce{CsPbI3} & 5.72 & $-$5.55 & $-$0.34 & $-$0.17 & 749 \cite{Sharma_Z_175_1992} & Y \\
    \ce{CH3NH3PbCl3} & 6.20 & $-$6.03 & $-$0.32 & $-$0.15 & $\cdots$ & Y \\
    \ce{CH3NH3PbBr3} & 5.80 & $-$5.81 & $-$0.11 & $-$0.11 & $\cdots$ & Y \\
    $\beta$-\ce{CH3NH3PbI3} & 5.66 & $-$5.61 & 0.01 & 0.06 & $\cdots$ & N \\
    \ce{CN3H6PbI3} & 5.43 & $-$5.43 & $-$0.38 & $-$0.39 & $\cdots$ & Y \\
    \ce{(CH3)4NPbI3} & 5.47 & $-$4.79 & $-$1.05 & $-$0.37 & $\cdots$ & Y \\
            \hline
    \ce{CsCl} & 2.59 & $-$6.64 & 0.14 & $-$3.91 & 918 \cite{Johnson_JACS_77_1955} & Y \\
    \ce{NaCl} & 3.01 & $-$8.22 & 1.46 & $-$3.76 & 1077 \cite{Hunter_PR_61_1942} & Y \\
        \end{tabular}
    \end{ruledtabular}
\end{table*}

\begin{figure}[h]
\centering
  \includegraphics[width=0.45\textwidth]{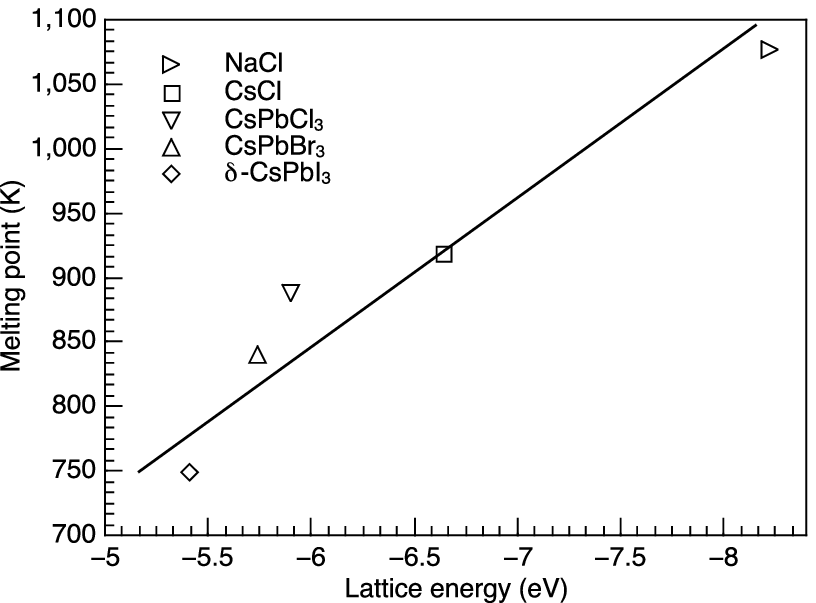}
  \caption{Correlation between the lattice energy and melting temperature of ionic compounds. The line is a guide to the eye.}
  \label{fgr:LEvsMP}
\end{figure}

The plot of the melting point \textit{vs} the lattice energy of those compounds is shown in Fig.~\ref{fgr:LEvsMP}.

From this figure, we can see that the melting point of different ionic structures including alkali halides follows a linear trend line. This suggests that formation of $A^+$ cations and  \ce{[BX3]-} complex anions is a result from melting of the perovskite structures.

\subsection{Stability analysis of hybrid organic halide perovksites}

Now we will utilize the Born-Haber cycle in order to evaluate components of the reaction enthalpy of hybrid halide perovskits. The lattice energies of \ce{CH3NH3PbCl3}, \ce{CH3NH3PbBr3} and  $\beta$-\ce{CH3NH3PbI3} perovskites are listed in Table~\ref{tbl:BH-perovskites}. All three compounds have similar values of the lattice energies ($\sim$10\% max-min difference). However, their stability characteristics are quite different. \citet{Buin_CM_27_2015} demonstrated that under ambient conditions \ce{CH3NH3PbCl3} and \ce{CH3NH3PbBr3} do not undergo a phase separation, unlike $\beta$-\ce{CH3NH3PbI3}. Both \ce{CH3NH3PbCl3} and \ce{CH3NH3PbBr3} remain stable up to the temperature of approximately 520~K, above which they decompose \cite{Nagabhushana_PNAS_113_2016}. Lattice energies of the corresponding inorganic perovskites (\ce{CsPbI3}, \ce{CsPbBr3} and \ce{CsPbCl3}) are very similar to their organic counterparts. In fact, these inorganic perovskites are chemically stable under the ambient environment. Remarkably, the lattice energy of $\beta$-\ce{CH3NH3Pbl3} and  $\delta$-\ce{CsPbCl3} structures are identical, in spite of the distinct stability characteristics. Therefore, we can conclude that the lattice energy cannot be used as a criteria to predict the chemical stability of compounds. 

The analysis of various contributions to the reaction enthalpies of hybrid halide perovskites (Table~\ref{tbl:BH-perovskites}) shows that the molecularization and lattice energies largely cancel each other. The ionization energy is the remaining contribution to the reaction enthalpy in Eq.~(\ref{Eq:BH_MAPbI3_dHr}) that ultimately controls the balance of the reaction. The lower $\Delta H_\text{ion}$ is, the more stable the compound.

Let us examine the chemical trends in ionization energy of various perovskites. The total ionization energy (Eq.~\ref{Eq:BH_MAPbI3_dHion}) comprises of two components: the ionization energy for the cation (\ce{Cs}$^+$ or \ce{[CH3NH3]+}) and that for the complex ion (\ce{[PbI3]-}, \ce{[PbBr3]-}, or \ce{[PbCl3]-}). Caesium has a lower ionization energy than \ce{CH3NH3} (Table~\ref{tbl:IE_OC}), which explains trends in the higher chemical stability of Cs-based perovskites as compared to their \ce{CH3NH3}-based counterparts. 

\begin{table}
  \caption{Ionization energies (eV) of atoms and molecules.}
  \label{tbl:IE_OC}
  \begin{ruledtabular}
  \begin{tabular}{lr}
    Ions & $\Delta H_\text{ion,1/2}$ \\
    \hline
    \ce{[(CH3)4N]+} & 3.15 \\
    \ce{Cs}$^+$ & 3.85 \\
    \ce{[CN3H6]+} & 3.81 \\
    \ce{[CH3NH3]+} & 4.20 \\
    \ce{[HCNH2NH2]+} & 4.30 \\
    \ce{[NH4]+} & 4.78 \\
    \ce{Na}$^+$ & 5.17 \\
    \ce{[CH3PH3]+} & 5.20 \\
    \ce{[CH3SH2]+} & 5.30 \\
    \ce{[PH4]+} & 5.36  \\
    \ce{[HCNH2PH2]+} & 8.36 \\
    \ce{[CH3]+} & 10.0 \\
        \hline
    \ce{[PbCl3]-} & $-$4.52 \\
    \ce{[PbBr3]-} & $-$4.31 \\
    \ce{[PbI3]-} & $-$4.19 \\
  \end{tabular}
  \end{ruledtabular}
\end{table}

Switching halides in the complex ions from \ce{PbI3} to \ce{PbCl3} lowers their electron affinity (Table~\ref{tbl:IE_OC}) and, thus, leads to the lower total ionization energy. This explains increase of the chemical stability when changing the inorganic cage from \ce{PbI3} to \ce{PbBr3} and \ce{PbCl3}.

In order to achieve a chemically stable hybrid halide perovskite structures, the necessary requirements are favourable geometrical factors (t-factor and octahedral factor) in conjunction with the low ionization energy ($\Delta H_\text{ion}\lesssim0$~eV).  Two strategies can be used to achieve this goal: (i) find a cation with the low ionization energy or (ii) select an inorganic cage with the low electron affinity. The second avenue is not very promissing, since the band gap of \ce{PbBr3}- and \ce{PbCl3}-based hybrid perovskites (2.3~eV \cite{Ryu_EEC_7_2014} and 2.9~eV \cite{Dimesso_CM_26_2014}, respectively) is outside of the favourable range for single-junction solar cells.

Since caesium has the lowest ionization energy in the periodic table, it is a challenging task to find molecules with smaller or similar ionization energy. Among the variety of organic cations listed in the Table~\ref{tbl:IE_OC}, \ce{[CN3H6]+} and \ce{[(CH3)4N]+} have the ionization energies lower than that for \ce{[CH3NH3]+} cation making them favourable candidates for perovskites with improved stability. However, the size of \ce{CN3H6} and \ce{(CH3)4N} molecules is significantly greater than \ce{CH3NH3}, which raises the tolerance factor above the upper formability limit of 0.95 (Table~\ref{tbl:IE_SOC}). 

\begin{table}
  \caption{Size of organic cations, the tolerance factor, volume of the unit cell and the band gap of selected perovskites.}
  \label{tbl:IE_SOC}
  \begin{ruledtabular}
  \begin{tabular}{lcccc}
    Perovskite & Cation radius  & Tolerance  & Volume & Bandgap \\
     & (pm)  & factor \cite{Kieslich_CS_5_2014, Liu_IC_55_2016} & ({\AA}$^3$/f.u.) & (eV) \\
    \hline
    $\beta$-\ce{CH3NH3PbI3} & 217 \cite{Kieslich_CS_5_2014, Liu_IC_55_2016} & 0.91 & 262 & 1.62 \\
    \ce{CN3H6PbI3} & 278 \cite{Kieslich_CS_5_2014, Liu_IC_55_2016} & 1.04 & 326 & 3.38 \\
    \ce{(CH3)4NPbI3} & 320 \cite{Palomo_JMS_215_2003, Garde_JCP_108_1998} & 1.15 & 361 & 3.30 \\
  \end{tabular}
  \end{ruledtabular}
\end{table}

From two structures,  \ce{CN3H6PbI3} and \ce{(CH3)4NPbI3} shows favourable reaction enthalpies of $-0.39$~eV and $-0.37$~eV, respectively(Table~\ref{tbl:BH-perovskites}). A large size of the organic molecule hinders formability of \ce{CN3H6PbI3} and \ce{(CH3)4NPbI3} perovksite structures. They both prefer hexagonal structures at ambient temperature \cite{Dimesso_MSEB_204_2016, Liu_IC_55_2016}. \citet{Marco_NL_16_2016} successfully synthesized and characterized \ce{CN3H6PbI3} perovskite solar cells. It was found that \ce{CN3H6PbI3} solar cell is also unstable under the ambient environment, which is evident from degradation of the power conversion efficiency over time. Interestingly, the rate of the efficiency decay is slower for \ce{CN3H6PbI3} as compared to \ce{CH3NH3PbI3}. \citet{SzafranskiTA_307_1997} found that \ce{CN3H6PbI3} crystals transform from orange-reddish phase to yellow phase after several hours at ambient pressure and temperature. This color changing demonstrates that the bandgap increases during phase transformation. From the reaction enthalpy of \ce{CN3H6PbI3} (Table~\ref{tbl:BH-perovskites}), we conclude that the drop of power conversion efficiency of \ce{CN3H6PbI3} photovoltaic device is due to the phase transformation, and the \ce{CN3H6PbI3} structure won't go through phase separation over time.

The ionization energies of onium ions in Table~\ref{tbl:IE_OC} correlate with the proton affinity of the corresponding molecules \cite{East_JACS_119_1997}. Molecules with the low ionization energy exhibit strong proton affinity and \textit{vice versa}. For instance, the proton affinity of \ce{PH3} is 785~kJ/mol, which is much lower than 901~kJ/mol for \ce{CH3NH2}. It turns out that methylamine has one of the strongest proton affinity among organic compounds. There very few organic molecules (including \ce{(CH3)2NH} studied here) with stronger proton affinity than \ce{CH3NH2}, but none of them have a size compatible with the \ce{PbI3} cage.

\section{Conclusions}
The Goldschmidt's tolerance and octahedral geometrical factors do not fully capture prerequisites for formability of hybrid halide perovskites. Here we used DFT calculations in conjunction with a Born-Haber cycle to evaluate contributions of the lattice, ionization and molecularization energies to the decomposition reaction enthalpy of hybrid halide perovskites. It was previously assumed that the instability of halide perovskite is due to a lower lattice energy of their ionic structure. We observe a correlation between the lattice energies and melting temperatures, but not with reaction enthalpies that are ultimately linked to the chemical instability of the perovskites. Analysis of Born-Haber cycle components suggests that the reaction enthalpy of hybrid halide perovskites is governed by the sum of ionization energies of a cation, e.g., \ce{[CH3NH3]+}, and an anion, e.g., \ce{[PbI3]-}. The lower total ionization energy, the more stable is the structure, provided the geometrical conditions are fulfilled (the tolerance and octahedral factors). This explains chemical trends in stability of hybrid and inorganic halide perovskites. For instance, the relatively high stability of \ce{CH3NH3PbCl3} is attributed to a lower ionization energy of \ce{[PbCl3]-} complex ion, whereas the stability of \ce{CsPbI3} is due to the lower ionization energy of \ce{Cs+}. The ionization energy of organic cations correlates with their proton affinity. In the search for hybrid perovskite with improved chemical stability and the band gap suitable for photovoltaic applications, several cations were investigated. The  promising candidates are \ce{[CN3H6]+} and \ce{[(CH3)4N]+} with the ionization energies even lower than \ce{Cs+}. The corresponding \ce{CN3H6PbI3} and \ce{(CH3)4NPbI3} structures have the decomposition reaction enthalpy approximately 0.3~eV more favourable than \ce{CH3NH3PbI3}. However, these ions has a prohibitively large size that translates into a large band gap. It is the fact that \ce{CH3NH2} has the highest proton affinity among molecules of comparable size. It makes challenging to find a cation suitable for \ce{PbI3} cage as a stable activate layer for photovoltaics.

%
%
\begin{acknowledgments}
Funding was provided by the Natural Sciences and Engineering Research Council of Canada under the Discovery Grant Program RGPIN-2015-04518. The work was performed using computational resources of the Thunder Bay Regional Research Institute, Lakehead University, and Compute Canada (Calcul Quebec).
\end{acknowledgments}


\end{document}